# ASAP-SML: An Antibody Sequence Analysis Pipeline Using Statistical Testing and Machine Learning

Xinmeng Li[1], James A. Van Deventer[2,3] and Soha Hassoun[1,2,*]

[1]Department of Computer Science, Tufts University, MA, United States of America

[2]Department of Chemical and Biological Engineering, Tufts University, MA, United States of America

[3]Department of Biomedical Engineering, Tufts University, MA, United States of America

\* Corresponding author

E-mail: Soha.Hassoun@tufts.edu

## Abstract

Antibodies are capable of potently and specifically binding individual antigens and, in some cases, disrupting their functions. The key challenge in generating antibody-based inhibitors is the lack of fundamental information relating sequences of antibodies to their unique properties as inhibitors. We develop a pipeline, Antibody Sequence Analysis Pipeline using Statistical testing and Machine Learning (ASAP-SML), to identify features that distinguish one set of antibody sequences from antibody sequences in a reference set. The pipeline extracts feature fingerprints from sequences. The fingerprints represent germline, CDR canonical structure, isoelectric point and frequent positional motifs. Machine learning and statistical significance testing techniques are applied to antibody sequences and extracted feature fingerprints to identify distinguishing feature values and combinations thereof. To demonstrate how it works, we applied the pipeline on sets of antibody sequences known to bind or inhibit the activities of matrix metalloproteinases (MMPs), a family of zinc-dependent enzymes that promote cancer progression and undesired inflammation under pathological conditions, against reference datasets that do not bind or inhibit MMPs. ASAP-SML identifies features and combinations of feature values found in the MMP-targeting sets that are distinct from those in the reference sets.

## Author summary

The availability of machine learning techniques and the exponential growth of sequencing data presents new opportunities to identify features that endow antibodies with the ability to disrupt the functions of biological targets. We have created a pipeline that uses statistical testing and machine learning techniques to determine features that are overrepresented in a specified set of antibody sequences in comparison to a reference set. The pipeline is referred to as Antibody Sequence Analysis Pipeline using Statistical testing and Machine Learning (ASAP-SML). We demonstrate the use of ASAP-SML by analyzing sets of antibodies that inhibit matrix metalloproteinases (MMPs) against reference sets. ASAP-SML performs within and across set similarity analysis. As in prior studies, our analysis of these datasets shows that features associated with the antibody heavy chain are more likely to differentiate MMP-targeting antibody sequences from reference antibody sequences. Further, ASAP-SML identifies several features in the MMP-targeting set that are distinct from the reference sets. Using design recommendation trees, ASAP-SML suggests combinations of features that can be included or excluded to augment the targeting set with additional candidate MMP-targeting antibody sequences.

## Introduction

Antibodies play an important role in treating diseases such as cancer and autoimmunity disorders by blocking specific protein-protein interactions and recruiting the immune system to specific cells and tissues. While experimental methods for antibody discovery, including hybridoma technology (1) and phage and yeast display (2), have allowed for significant advances in discovering specific binding proteins, difficulties remain in establishing general strategies for designing antibodies that disrupt enzymatic activity or other biological functions. In particular, relating the amino acid sequences of these antibodies to their unique abilities in disrupting biological functions remains a challenge. Data-driven computational approaches may shed light on such fundamental information. Recent computational tools provide first steps towards elucidating structural information that can guide rational antibody design. Several numbering tools (e.g., AbNum (3), DomainGapAlign (4), PyIgClassify (5), ANARCI (6), and AbYsis (7)) annotate an antibody sequence to identify the Complementary Determining Regions (CDRs) and the Framework Regions (FRs). There are also tools (e.g., IgBLAST (8) and SAbDab (9)) to select templates from databases for the variable domains VH and VL. Other tools (e.g., PIGS (10), FREAD (11), PyIGClassify) predict the structures of CDR loops. Efforts to partially design antibody sequences that bind to specific targets have been made utilizing several computational tools (e.g. OptMaven (12) and RAbD (13)). These tools graft designed CDRs on backbones, and then utilize energy minimization and optional docking procedures to obtain complete antibody sequences. Despite these and other efforts (14, 15), there remains a critical gap in linking antibody sequences directly to biological consequences such as target inhibition.

We address in this paper the challenge of identifying features of antibodies that may influence antibody function. We design a pipeline for analyzing antibody sequences and extracting *features* (e.g., germline, positional motifs, etc.) and *feature values*

(e.g., the specific sequence of residues in the CDR-H3 region) that are overrepresented in one dataset, referred to here as a targeting set, as compared to a reference dataset. Our approach is data-driven, enabled by the increasing availability of amino acid sequences of functional antibodies in databases (e.g. Protein Data Bank (PDB) (16), IMGT (17)) and in patents. The pipeline is termed Antibody Sequence Analysis Pipeline using Statistical testing and Machine Learning (ASAP-SML). The pipeline extracts features associated with each antibody sequence, as well as features specific to the CDR-H3 region due to its role as the primary specificity determinant of most antibodies (18, 19). ASAP-SML then utilizes several machine learning techniques and statistical testing to determine important and statistically significant features that distinguish the sequences of targeting antibodies from the sequences within the reference antibodies and, if appropriate, to recommend combinations of design features that can be utilized during efforts to improve the binding properties of existing antibodies or to search for new antibodies that target an antigen of interest.

The pipeline can be applied to contrast any two antibody data sets. For example, the targeting set may contain antibodies that interfere with their biological targets, collected through experiments, patents and/or database searches, while the reference set may contain antibody sequences that are curated from the Protein Data Bank and other reference sources. In this context, and for large and diverse target and reference sets, ASAP-SML operates to identify features that are overrepresented or underrepresented within the target set in comparison to the reference set. To demonstrate the use of ASAP-SML, we apply it to analyze eight datasets of antibodies that inhibit matrix metalloproteinases (MMPs). The MMPs are a family of zinc-dependent enzymes that play numerous roles in normal physiology and development, but under pathological conditions, dysregulated MMPs can facilitate cancer progression (20, 21), undesired inflammation (22), and other conditions. Among hundreds of features, ASAP-SML identified several salient feature values for the MMP-targeting antibody data sets that distinguish it from reference data sets. We note that analysis via our pipeline may be confounded by limited data availability, or differences in antibody sequences that arise for reasons outside of function disruption (e.g., the specific antibody libraries, or screening procedures used to isolate antibodies of interest).

## Methods

### ASAP-SML overview

The ASAP-SML pipeline comprises five steps (Figure 1). In the Data Preparation step, amino acid sequences for both targeting antibodies and reference sequences are prepared for use within the pipeline. In the Sequence Numbering step, each antibody sequence is annotated with a numbering scheme to allow for the identification of the six Complementary Determining Regions (CDRs), three on the heavy chain (H1-H3) and three on the light chain (L1-L3)) directly involved in antigen binding (23), and the remaining Framework Regions (FR). In the Feature Extraction step, features associated with each antibody sequence (predicted germline and canonical structures of CDRs) and features specific to the CDR-H3 region (Isoelectric point (pI) and frequent positional motifs) are extracted. In the Analysis step, salient features are identified using statistical testing and machine learning techniques. Finally, in the Design Recommendation step, decision trees (24) are used to identify combinations of salient feature values for inclusion in targeting antibody sequence design. The ASAP-SML pipeline was released as an open-source python code. The pipeline and all datasets are available on GitHub (https://github.com/HassounLab/ASAP-SML).

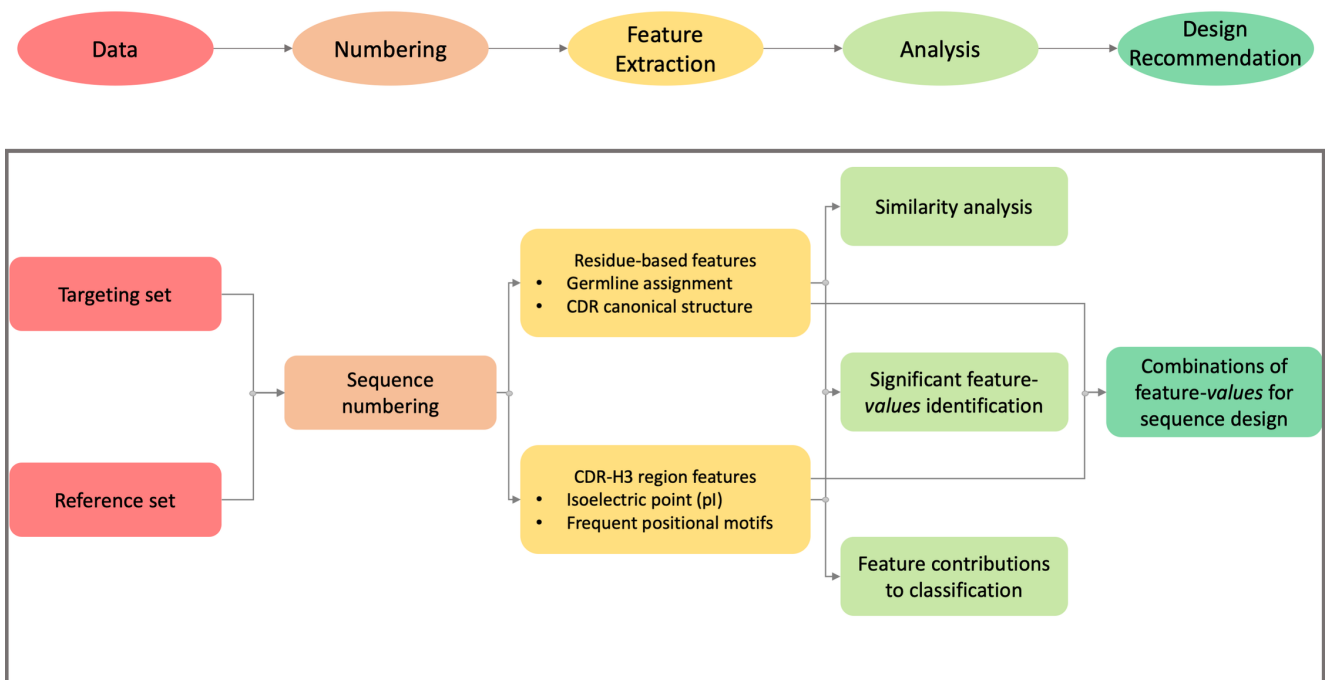

**Figure 1. ASAP-SML pipeline overview.** Antibody sequences in the targeting and reference sets are inputted into the pipeline to perform sequence numbering, feature extraction, sequence and feature analysis, and design recommendations.

## Data, and data preparation

The user inputs two data sets, a *targeting set* and a *reference set*. The user may partition sequences within each set into groupings, each referred to as a *dataset*. Such datasets, for example, may be collected through different experiments or data sources. ASAP-SML accommodates various size sets and datasets. The numbering and feature extraction steps of our pipeline analyze every sequence within each set. In the analysis and design recommendation steps, sequences are sampled, and analysis results are aggregated across a large number of sampling iterations, $k$. We assume a default iteration number $k = 100$. To facilitate sampling, the median size among the datasets within the targeting set is selected as a *desired dataset size.* Each dataset is then either sampled or duplicated to achieve the desired dataset size. When sampling is required, the number of sampled sequences in the targeting set is set to the product of the desired dataset size times the number of datasets. This size is also used for the number of desired sampled sequences in the reference set and drives the numbers of samples per dataset in the reference set. The sampling parameters for the targeting and reference datasets can be overridden by the user.

**Antibody sequence numbering**

Aligning antibody sequences to a consensus sequence through a numbering scheme enables dividing an antibody sequence into six Complementary Determining Regions (CDRs), which are directly involved in antigen binding, and Framework Regions (FRs). We elected to use the Chothia numbering scheme (25) rather than other common numbering schemes, e.g., IMGT (26), Kabat (27), Martin (3), and AHo (28). This decision facilitated further downstream analysis when predicting CDR-H3 canonical structures using PIGS, which depends on the Chothia numbering scheme. ANARCI (6) is used to assign the Chothia numbering scheme to each antibody sequence in the targeting and reference sets.

**Feature extraction**

We extract four types of features in this step, germline, CDR canonical structure, pI (isoelectric point) range, and frequent positional motifs in CDR-H3. For each sequence, values for each feature are determined and recorded in a vector we refer to as a feature fingerprint. Each entry in the fingerprint is assigned either a "1" or "0," indicating the presence or absence of a particular feature value within the antibody sequence. The number of possible feature values, and therefore the width of the fingerprint, is dependent on the sequences in the reference and targeting sets.

Germlines are encoded in immature B cells and are used as templates for generating diversity during selection of antibodies against specific targets (25). Putative germlines for each heavy and light antibody sequences on the V and J regions (HV, HJ, LV, and LJ) are assigned using ANARCI. ANARCI performs an alignment between the query antibody sequence and multiple pre-built Hidden Markov Models (HMMs). The putative germline is determined for an antibody sequence by the most significant alignment among all HMMs for a domain type of a particular species (Human, Mouse, Rat, Rabbit, Pig or Rhesus Monkey). ASAP-SML allows germline assignments from any of the species included in ANARCI. As some allelic variations involve a single base change, they were ignored for all germline assignments.

Structural conformations for the highly variable CDRs are known as canonical structures. These structures can be predicted based on both loop length and amino acid identities at specific position (25). Here, we use the canonical structure determination rules introduced in the Prediction of ImmunoGlobulin Structure (PIGS) database (10) to determine if a CDR canonical structure assignment can be made for a given sequence, and, if so, the specific canonical structure is assigned.

Due to the important role of the CDR-H3 region in antibody-antigen interaction specificity, two types of features, pI and frequent positional motifs, are generated to characterize antibody sequences. The isoelectric point (pI) is defined as the pH value when net positive and negative charges of an amino acid sequence are balanced. pI values impact protein properties including solubility and stability (29). pI values of CDR-H3 sequences are calculated using *bio.sequtils.isoelectricpoint* from the *biopython* package (30). PI values are binned with a data-driven method. The range [0, 14] is initially divided into two equal sized bins, and then recursively halved. Bin halving is terminated if a bin contains less than 10% of antibody sequences or if the bin size reaches a range of pI values equal to or less than 0.3. This range is assumed as the minimum range that provides specificity differentiation.

We define motifs as sequential amino acid sequences ranging from 2 to 10 amino acids in length within the CDR-H3 region. We consider motifs along with their positional information. The starting point of a positional motif is defined using the position of the first amino acid within the motif. Each positional motif is labeled with its starting point, an underscore, and the amino acid sequence. For example, positional motif "4_AL" indicates that there is a motif of length two consisting of sequence "AL" that starts at position four of the CDR-H3 region. For each antibody sequence in the targeting and reference subsets, the two most frequent positional motifs of each length in the CDR-H3 region are included in the feature fingerprint, thus allowing for a variety of subsequence lengths for the positional motifs.

## Sequence and feature analysis

Several analysis methods are used to contrast the sequence and feature fingerprints for the targeting and non-targeting sets. First, pairwise similarity is computed based on both sequence similarity and also on fingerprint similarity for each pair of sequences. Pairwise similarity of the heavy chain and light chain sequences is calculated for the Chothia-numbered sequences. When comparing two sequences, the similarity of amino acids in each position is looked up in the BLOSUM62 matrix, which provides a match/mismatch score between two amino acids in reference to protein sequences in the BLOCKS database (31). An amino-acid insertion between two numbered positions or a deletion at a numbered position is considered as a gap at the position in question. Default gap penalties are applied. The match/mismatch scores are summed and rescaled based on min-max normalization (32). The feature rescaling step computes a score that is first adjusted by the minimum set score over the set of match/mismatch scores for each chain, and then normalized by dividing the adjusted score by the difference between the minimum and maximum set scores. The rescaling step thus eliminates contributions to the similarity scores due to constant region sequences within the light chains and within the heavy chains. The rescaled normalized score is used as the pairwise sequence similarity score.

When comparing feature fingerprints, pairwise similarity is computed per fingerprint segment corresponding to each feature (pI and frequent positional motifs) or feature region (germline-HV, germline-HJ, germline-LV, germline-LJ, CDR-H1, CDR-H2, CDR-H3, CDR-L1, CDR-L2, and CDR-L3). For each segment, the similarity score is computed using the Jaccard index, which computes the size of the common features divided by the number of features present in either feature fingerprint. Each segment score is given a weight of 1. The weighted sum of the segment scores, normalized to the maximum possible score, provides the pairwise fingerprint similarity score.

Pairwise similarity scores are visually inspected using heat maps. Statistical testing is then used to quantify sequence and feature similarity trends. Our first statistical test examines the within-set similarity in the targeting set against the within-set similarity for the non-targeting set. Our null hypothesis assumes that the pairwise similarity scores in the targeting set have the same statistical distribution as those in the reference set. Our second test examines how the extracted features correlate with the heavy-chain antibody sequences and with the light- chain antibody sequences. Our null hypothesis assumes that there are no differences in how the extracted features correlated with the heavy and light chain sequences. A one-tailed Wilcoxon rank-sum test from the *scipy.stats.mannwhitneyu* python package is used to perform these statistical tests. The test is repeated for k iterations, where representative datasets are created each time through either sampling or duplication. The test is significant if all iterations report a *p*-value less than 0.05.

The second analysis method identifies salient feature values that differentiate the targeting and reference sets. Statistical testing for each feature value for the features in Table 1 using Fisher Exact Test (FET) identifies statistically significant feature values with *p*-values less than 0.05. Additionally, feature selection is performed using a random forest algorithm (*sklearn.ensemble* python package), and importance scores are calculated. FET analysis and random forest analysis are repeated for k iterations. The reported *p*-value and rank for each feature are averaged across k iterations. Unlike FET, a frequentist approach, feature selection evaluates the contribution of feature values to classification. Therefore, the two tests provide complementary analysis.

While salient feature analysis identifies feature values that differ between sets, a third analysis method evaluates the contributions of features or combination of features in classifying MMP-targeting and non-targeting sequences. We evaluate the contributions using all features or using only one type of feature. To ensure that the results are consistent regardless of the classification method, three classification techniques are used: Support Vector Machine (SVM) (33), random forest (34) and Adaptive Boosting (AdaBoost) algorithms (35). The classification accuracy is measured using AUC, the area under the

**Table 1. Extracted features**. Listing of (a) features in the fingerprint vector, (b) regions within antibody that exhibit the feature, (c) software extraction method, and (d) number of possible feature values for the MMP-targeting set test case.

| Feature (a) | Region (b) | Extraction method (c) | Number of possible feature values for MMP-targeting set (d) |
|---|---|---|---|
| Germline | HV | ANARCI | 110 |
| | HJ | | 6 |
| | LV | | 99 |
| | LJ | | 11 |
| CDR Canonical Structure | H1 | PIGS | 5 |
| | H2 | | 5 |
| | H3 | | 4 |
| | L1 | | 4 |
| | L2 | | 1 |
| | L3 | | 1 |
| Isoelectric Point (pI) (CDR-H3) | | *biopython* package | 8 |
| Frequent Positional Motifs (CDR-H3) | | ASAP-SML script | 46 |

Receiver Operating Characteristic (ROC) curve, based on ten-fold cross-validation. Therefore, the set of sequences is randomly partitioned into 10 equal-size subsamples. Of the 10 subsamples, a single subsample is used for validation while the remaining 9 subsamples are used as training data. To achieve ten-fold validation, this process is repeated 10 times, where a different subsample is selected each time for validation. Further, the AUC is averaged across *k* sampling iterations.

**Design recommendation using design recommendation trees**

The prior analysis step identifies salient features in the targeting antibody set. The design recommendation step identifies combinations of presence/absence of such feature values that are distinct within each set. To this end, ASAP-SML learns a decision tree, a flow-chart like structure that segments the data into sets that have particular features. The decision tree therefore identifies combinations of non-conflicting salient features that are distinct when comparing the targeting dataset with the reference dataset. Further, we augment the decision tree to show the percentage of existing sequences with such combination features in both reference and targeting set. With the exception of leaf nodes, each node in the tree tests the presence or absence of a particular feature value. The feature value with the lowest misclassification rate, the Gini impurity (36), is used to partition the node into two child nodes. Left branches represent the true outcome of the test, while the right branch represents the false outcome. Decision tree branching is stopped when at least one leaf has less than 5% of the total number of antibody sequences associated with the root node.

The decision tree algorithm outputs a colored-tree diagram (Figure S3). Each node in the decision tree is colored based on the ratio of its number of sequences from each targeting and reference sets. Nodes with more targeting antibody sequences are colored in blue, and nodes with more reference antibody sequences are colored in orange. A path from the root node to a particular tree node represents a combination of feature values that are either mostly excluded or included for the sequences associated with the tree node. The true branches along such a path thus correspond to feature values that are present in the leaf-node sequences, while false branches correspond to excluded feature values. The *sklearn.tree.DecisionTreeClassifier* (37) python package is used to construct the decision tree.

While the decision tree identifies a model that best classifies the sequences, our intention is to utilize the decision tree to identify combinations of feature values to use when designing targeting sequences. We therefore augmented the tree with two new metrics, split efficiency (SE) and error rate (ER). Split efficiency is calculated as the number of targeting antibody sequences in the current node divided by the number of targeting antibody sequences in the root node. Split efficiency reflects the portion of sequences in targeting set that has the combination of valid feature value following the path from root to current node. The split efficiency of a node is less than or equal to that of its parent node. Error rate is calculated as the number of reference antibody sequences in the current node divided by the number of antibody sequences from both the targeting and reference sets in the current node. Error rate tells the likelihood of having a non-targeting sequence when including the design features from the root to the current node. Error rates are independent of tree depth (distance from root). To identify the best combination of design features, all paths from root to nodes that are dominated by the targeting sequences are analyzed with the goal of identifying node(s) that maximizes split efficiency and minimizes error rate. We refer to this newly labeled decision tree as a *design recommendation tree*.

## Results

### MMP-targeting and reference data

The data used to illustrate the functionality of ASAP-SML is composed of publicly available amino acid sequences of antibodies reported to inhibit MMPs and two reference datasets. A summary of the MMP-targeting antibody datasets is listed in Table 2, ordered in decreasing number of antibody sequences available in each dataset. Seven of the eight datasets were collected from patents in which experimental data is presented confirming the inhibitory activity of at least a portion of the antibody sequences that bind to a member of the human MMP family. We note that MMP-targeting datasets 1-5 all originated from work conducted at a single company (Dyax) and these datasets contain sequences almost exclusively based on the IGHV3-23 germline (38). An additional dataset, labeled MMP-targeting 8, was collected from the PDB by searching for MMP-targeting antibodies using the keywords "MMP" and "antibody". The identified antibody sequences were deposited by Udi et al (39). The sizes of the MMP targeting datasets are highly variable. The largest dataset, MMP-targeting set 1, has 621 antibody sequences, while MMP-targeting datasets 7 and 8 only have 4 antibody sequences each. Antibody sequences collected here target MMP-2, -9, -12, -13, -14, -26. MMP-targeting datasets 2 and 5 are culled from the same patent (40), where sequences in dataset 5 inhibit MMPs, while sequences in dataset 2 are known to only bind to MMPs.

**Table 2. The MMP-targeting antibody set comprises 8 datasets.**

| MMP-targeting Dataset Number | Targeted MMPs | Sequences source | Reference | Number of sequences | Number of representative sequences |
|---|---|---|---|---|---|
| 1 | MMP-2, MMP-9 | US8013125B2 | (41) | 621 | 16 |
| 2 | MMP-12 | US8114968 | (40) | 69 | 64 |
| 3 | MMP-26 | US20060036076A1 | (42) | 44 | 43 |
| 4 | MMP-14 | US7745587B2 | (43) | 12 | 12 |
| 5 | MMP-12 | US8114968 | (40) | 12 | 12 |
| 6 | MMP-13 | WO2007065037A2 | (44) | 11 | 11 |
| 7 | MMP-9 | US8377443 | (45) | 4 | 1 |
| 8 | MMP-14 | Protein Data Bank | (39) | 4 | 1 |
| Total Number of sequences | | | | 777 | 160 |

Upon initial analysis, there were 24 antibody sequences in the MMP-targeting set that were assigned by ANARCI to non-human or to non-murine germlines. These sequences were removed from the MMP-targeting set. Additionally, high sequence similarity was observed within some of the individual datasets due to the presence of antibody sequences from *in vitro* affinity maturation campaigns in which point mutants of a single parent antibody were evaluated for improved affinity. BLAST-CLUST (46) was used to cluster highly similar sequences (up to 95% similarity in residues after sequence alignment), and to select a representative sequence for each cluster. Clustering revealed that the largest dataset, MMP-targeting dataset 1, contains only 16 distinct clusters from the 621 antibody sequences extracted from the US8013125B2 patent. MMP-targeting dataset

2, which includes 69 antibody sequences extracted from patent US8114968, still contained 64 distinct sequences after clustering, MMP-targeting dataset 3 contains 43 distinct sequences after clustering. A total of 160 representative sequences were merged into a combined MMP-targeting set for further analysis. This representative set is referred to at the *MMP-targeting set*.

To identify a set of antibody sequences that do not bind to or inhibit MMPs, the PDB was queried on May 24, 2017 for human and murine sequences that do not bind to or inhibit MMPs. The inclusion of both human and murine sequences in this reference set was deliberate, as some of the MMP-targeting antibodies present in the datasets are murine in origin due to the immunization-based strategies used to generate the sequences. To avoid overrepresentation of highly similar antibody sequences in the reference dataset, the database was queried for representative sequences with 95% or fewer identical residues. Only antibody sequences with paired heavy and light chains and complete variable regions were selected. Sequences that were assigned by ANARCI to non-human or to non-murine germlines were excluded. The resulting reference dataset consisted of 183 human antibody sequences and 197 murine antibody sequences. This reference dataset is referred to as the *PDB-reference set*.

When analyzing the MMP-targeting set against the PDB-reference set, we observed that 92.50% of the MMP-targeting set sequences were assigned to IGHV3-23, consistent with the origination of MMP-targeting sets 1-5 from Dyax (38). This feature is therefore not likely to be functionally significant. To remove the potentially confounding effects of this germline overrepresentation, we selected a subset of the MMP-targeting set that contains only the subset of sequences with the IGHV3-23 germline and then used BLASTCLUST to select representative sequences. This reduced set is referred to as the *MMP-IGHV-targeting set*. Instead of comparing this set against the PDB-reference set, which contains a large variety of germlines, we compared the MMP-IGHV-targeting set against IGHV3-23 human antibody heavy-chain sequences from former studies (European Molecular Biology Laboratory accession numbers AM076988–AM083316) (47-50). These sequences, which were utilized to study CDR diversity within a controlled germline context, are not known to bind to any specific targets. This reference set is referred to as the *IGHV-reference set.* As only the heavy chain sequences were available for the IGHV-reference set, ASAP-SML was applied to compare the MMP-IGHV-targeting and the IGHV-reference sets. MMP-IGHV-targeting set has 134 representative sequences and the IGHV-reference dataset has 4673 representative sequences.

We applied ASAP-SML to these sets. For each set, sequences were numbered and then extracted. The analysis is reported below first for the MMP-targeting against the PDB-reference set, followed by the MMP-IGHV-targeting against the IGHV-reference set. The design recommendation step is only demonstrated for the MMP-targeting against PDB-reference set. The MMP set described in Table 1, its features, and the mapping from the original set to the MMP-targeting set and the MMP-

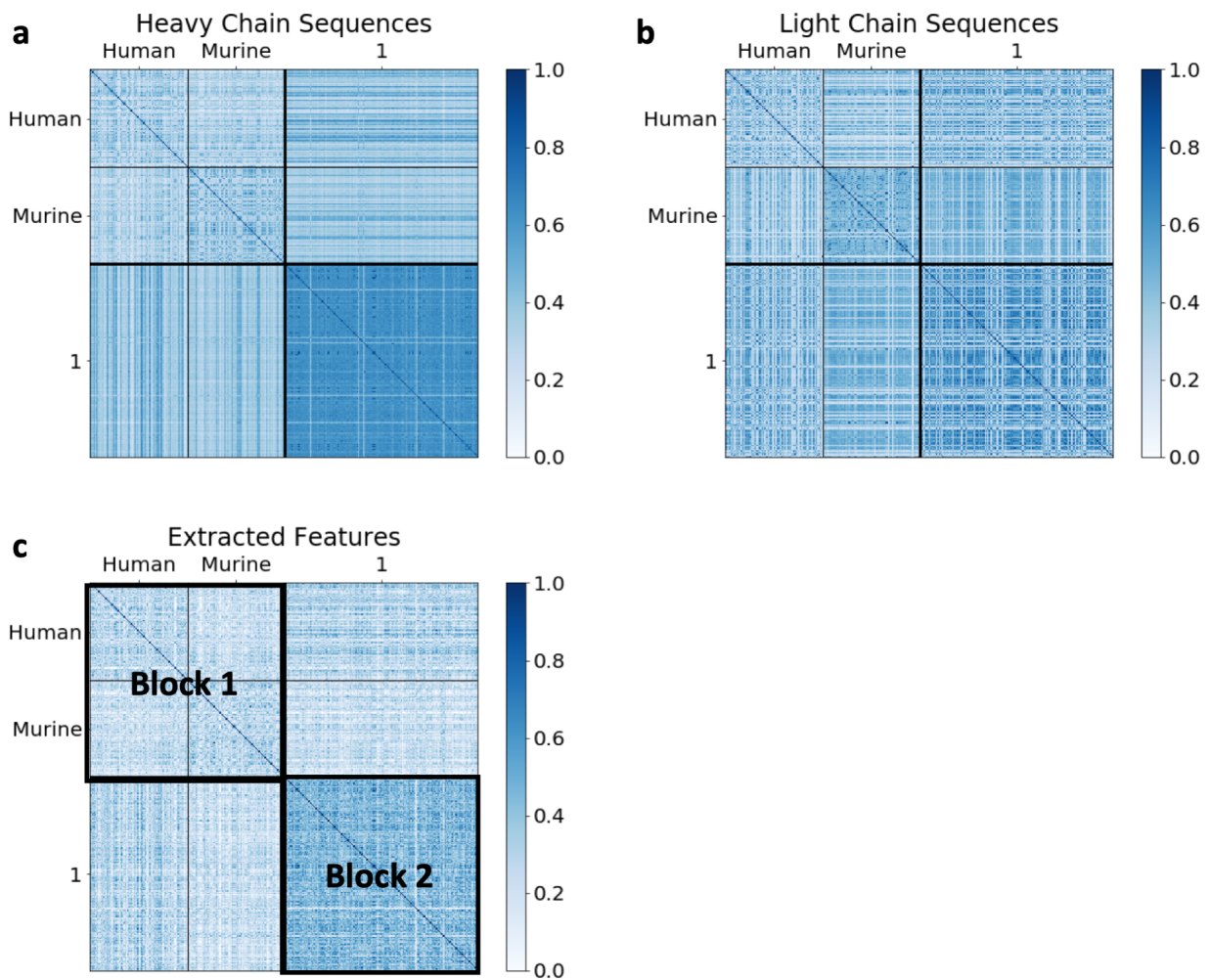

**Figure 2. Heat maps comparing the reference set, consisting of human and murine antibody datasets, with the MMP-targeting set, consisting of datasets 1**-8. (a) Heavy-chain sequence similarity heat map, (b) Light-chain sequence similarity heat map, (c) Extracted-feature similarity heat map. To visualize within-set similarity for the reference set and within-set similarity for the MMP-targeting set, the sets are marked with Block 1 and Block 2, respectively, on the extracted-feature heat map.

IGHV-targeting set, are provided in Table S1. Sequence details and analysis of the MMP-targeting set against the PDB-reference sets are provided in Table S2. Sequence details and analysis of the MMP-IGHV-targeting set against the IGHV-reference set are provided in Table S3.

### Similarity analysis for MMP-Targeting vs PDB-reference sets

**Heat maps.** To analyze similarities within and between MMP-targeting vs PDB-reference sets, heat maps were constructed using heavy-chain pairwise sequence similarity (Figure 2a), light-chain pairwise sequence similarity (Figure 2b), and *feature value* pairwise similarity (Figure 2c). For all heat maps, datasets within the MMP-targeting set exhibit high pairwise similarity, despite the fact that the datasets include sequences that target a variety of MMPs (e.g., MMP-2, MMP-9, etc.). This trend appears stronger in the heat map for feature values (Figure 2c), but less so in the heat map for the light-chain sequences (Figure 2b). Further, heavy-chain pairwise similarity tracks the extracted-feature similarity more so than the light-chain pairwise similarity.

**Statistical testing.** ASAP-SML provided a statistical test to confirm that within-set similarity in the MMP-targeting set is higher than within-set similarity for the reference set. The testing was performed for the targeting pairwise similarity data for heavy-chain sequences, light-chain sequences, and extracted features, against their respective pairwise similarity data of the reference set, ASAP-SML performed a one-tailed Wilcoxon rank-sum test to evaluate the similarities within sets. The test was repeated k = 100 times. The average *p*-values for each of the three heat maps was less than 0.001, indicating statistical significance. Therefore, pairwise within-set similarity scores for the MMP-targeting set are higher than those for the reference set, confirming the observation that the relationships among the set of antibody sequences in the MMP-targeting set are closer than those sequences found in the reference set.

ASAP-SML also provided a statistical test to confirm that the extracted features better correlate with the heavy chain antibody sequences than with the light chain antibody sequences, as indicated by visual inspection of the heat maps. Differences between heavy-chain pairwise sequence similarity and feature-value pairwise similarity, as computed for Figures 2a and 2c, and differences between light-chain pairwise sequence similarity and feature-value pairwise similarity, as computed for Figures 2b and 2c were computed. Our null hypothesis assumes that these two computed differences have the same statistical distribution. A one-sided Wilcoxon rank-sum test was then performed on the differences to evaluate the alternate hypothesis that the computed differences between heavy-chain pairwise sequence similarity and feature-value pairwise similarity is more significant than the computed differences between light-chain pairwise sequence similarity and feature-value pairwise similarity. The test was performed for $k=100$ repetitions. The *p*-value was less than 0.001 for each repetition. Therefore, the extracted-feature heat map (Figure 2c) is more correlated with the heavy-chain heat map (Figure 2a) than with light-chain heat map (Figure 2b). With the limited sequence diversity present within the targeting set, the observed correlations may be artifactual in nature. Larger and more diverse collections of targeting sets would be needed to fully validate the correlations observed in our analyses.

**Table 3.** Top 5 salient feature values as determined by Fisher Exact Test.

| Rank | Feature | Feature value | Average *p*-value | Frequency in PDB-reference set | Frequency in MMP-targeting set |
|---|---|---|---|---|---|
| 1 | Germline HV | IGHV3-23 | 4.51E-60 | 5.26% | 92.50% |
| 2 | CDR Canonical Structure H2 | Type 6 | 4.02E-34 | 27.11% | 94.38% |
| 3 | Germline LJ | IGKJ3 | 2.64E-05 | 2.11% | 16.25% |
| 4 | Germline HJ | IGHJ6 | 1.64E-04 | 16.05% | 36.25% |
| 5 | Germline LV | IGKV1-39 | 1.74E-04 | 3.95% | 17.50% |

**Table 4.** Top 5 salient feature values as determined by feature selection.

| Rank | Feature | Feature value | Importance score | Frequency in PDB-reference set | Frequency in MMP-targeting set |
|---|---|---|---|---|---|
| 1 | Germline HV | IGHV3-23 | 0.2605 | 5.26% | 92.50% |
| 2 | CDR Canonical Structure H2 | Type 6 | 0.1084 | 27.11% | 94.38% |
| 3 | CDR Canonical Structure H2 | Type 5 | 0.0699 | 39.47% | 0.00% |
| 4 | Germline HJ | IGHJ2 | 0.0266 | 1.25% | 18.95% |
| 5 | Germline LJ | IGKJ3 | 0.0150 | 2.11% | 16.25% |

**Salient feature-value identification for MMP-Targeting vs PDB-reference sets**

ASAP-SML sought to identify individual extracted feature values associated with the MMP-targeting set (Table S2f). For the 300 extracted features, the FET identified 35 significant features, while the random forest model for feature selection identified 60 features as important. Of these feature values, 26 were identified by both methods. Out of these 26 features, 8 germlines, 5 CDR canonical structures, 2 pI ranges and 11 frequent positional motifs were identified. The FET identified 18 of the 184 possible heavy chain features as significant, while only 7 of 116 possible light chain features were identified as such.

To analyze frequencies of salient feature values, frequency analysis is applied to the MMP-targeting and PDB-reference sets (Table S2e). The frequency of the most "important" distinguishing feature based on both FET and importance analysis, germline IGHV3-23, is 92.50% in the MMP-targeting sequences but only 5.26% in the reference dataset sequences. This is likely the result of the datasets used as inputs for the pipeline in this work and may not be directly attributable to functional differences. The frequency of the second most important feature, CDR-H2 canonical structure type 6, is 94.38% in the MMP-targeting sequences and 27.11% in the reference sequences. The differences in frequency of the remaining top 5 features (Tables 3 and 4) were less than ~40%. The absolute differences in the frequency of feature values in the reference and MMP-targeting set are reported (Table S2e). A feature value with high differences in frequency (> 50%) is considered a *biasing feature*, a feature that can distinguish two sets with high classification accuracy. Germline IGHV3-23 and CDR-H2 canonical structure type 6 are the two biasing features when analyzing the MMP-targeting versus PDB-reference sets. We emphasize that these features are biasing *within* the context of our datasets. However, these features may not be biasing when considering a different MMP-targeting dataset and/or different reference sets.

To explore if any of the salient features are correlated, associations amongst pairs of features from the MMP-targeting and the PDB-reference sets were computed (Table S2g). The Jaccard coefficient is used to compute the co-occurrence of each pair of binary (0 or 1 depending on feature presence or absence) features (51). Given two feature vectors for sequences within a particular dataset, the Jaccard coefficient represents the proportion of sequences that have both features present relative to the total number of sequences where at least one of the two features are present. A high Jaccard coefficient score (> 0.8) suggests strong association relationship between two feature values. Within the MMP-targeting set, the two biasing features identified in previous frequency analysis, germline IGHV3-23 and CDR-H2 canonical structure type 6, have strong association with a Jaccard coefficient score of 0.90. Additionally, CDR-H1 canonical structure type 1, CDR-L2 canonical structure type 0 and CDR-L3 canonical structure type 0 have strong association with each of the two biasing features. There were also highly associated features within the PDB-reference set (Table S2g). All of these observations confirm that our pipeline is able to identify features that are distinct between two datasets.

**Contribution of features to classification for MMP-Targeting vs PDB-reference sets**

To assess if extracted features can distinguish targeting and reference sequences, ASAP-SML analyzed the performance of three classification algorithms, SVM, random forest, and AdaBoost, in separating MMP-targeting from reference antibody sequences using all features or based on one type of feature only. To assess the impact of the biasing features, classification was re-run with the exclusion of biasing features and their associated features. In both cases, AUC data using all three algorithms yielded similar ROCs. The following discussion explicitly refers to the results based on the SVM AUC data, but it is generalizable to other algorithms.

Classification using all features with SVM yielded an AUC of 0.9812 (Figure S1a). Classification was re-run while retaining one feature-at-a-time (Figure S1b-e) yielding the following AUC values: germline, AUC=0.9750; CDR canonical structures,

AUC=0.8414; pI, AUC=0.6383, and frequent positional motifs, AUC=0.6960. Due to the biasing features, the AUC was high when using germline and CDR canonical structure features.

We further investigated classification when including subsets of features (Figure 3). The biasing features are germline IGHV3-23 and CDR-H2 canonical structure type 6, while CDR-H1 canonical structure type 1, CDR-L2 canonical structure type 0 and CDR-L3 canonical structure type 0 are deemed associated features. Excluding biasing features and their associated features and using the remaining features that are included in Table S2e, SVM classification yielded an AUC of 0.8668 (Figure 3a). Classification while retaining one feature-at-a-time (Figure 3b-e) yielded a range of AUC values: germline, AUC=0.8396; CDR-canonical structure, AUC=0.6248; pI, AUC=0.6383, and frequent positional motifs, AUC=0.6960. Despite excluding the biasing germline (IGHV3-23), the germline AUC was surprisingly high. Examining the frequencies of the non-excluded germlines, HJ, LV, and LJ, showed that the germlines of the two datasets were mutually exclusive (Table S2e), thus providing two groupings of biasing germline features. The AUC for the CDR canonical structure was significantly lower when excluding the biasing features due to the removal of CDR-H2 canonical structure type 6 and other associated CDR canonical structure features (CDR-H1, CDR-L2 and CDR-L3) as classification features. The AUC for pI and the frequent positional motifs were identical to those in the prior classification case. We further investigated the impact of removing all germline features and associated CDR canonical structure features and retaining the remaining CDR canonical structure features, pI features and frequent positional motifs features. Classification yielded an AUC of 0.7599 (Figure 3f). The group of features utilized in Figure 3f proved an important orthogonal predictor for classification. Further, many such features were identified earlier as important using FET and importance feature selection. These findings suggest that combinations of extracted features can distinguish the sequences in the MMP-targeting vs PDB-reference sets. For more large and diverse datasets than what we have analyzed herein, this classification procedure can highlight features for further biological characterization to determine their exact roles in binding to targets and disrupting biological function.

**Salient feature-value identification for MMP-IGHV-targeting and IGHV-reference sets**

For the 57 extracted features, FET identified 17 significant features, while the random forest model for feature selection identified 19 features as important; 7 of these feature values were identified in both methods (Table S3f). Out of these 7 features, 1 germline, 1 CDR canonical structures, 1 pI ranges, and 4 frequent positional motifs were identified. Frequency analysis of features within the MMP-IGHV-targeting and IGHV-reference sets (Table S3e) showed that none of the salient features identified in FET or feature selection have high frequency (> 80%) or high difference in frequency between the two sets (>50%). This result suggests that there are no biasing features when analyzing the MMP-IGHV-targeting and IGHV-reference sets.

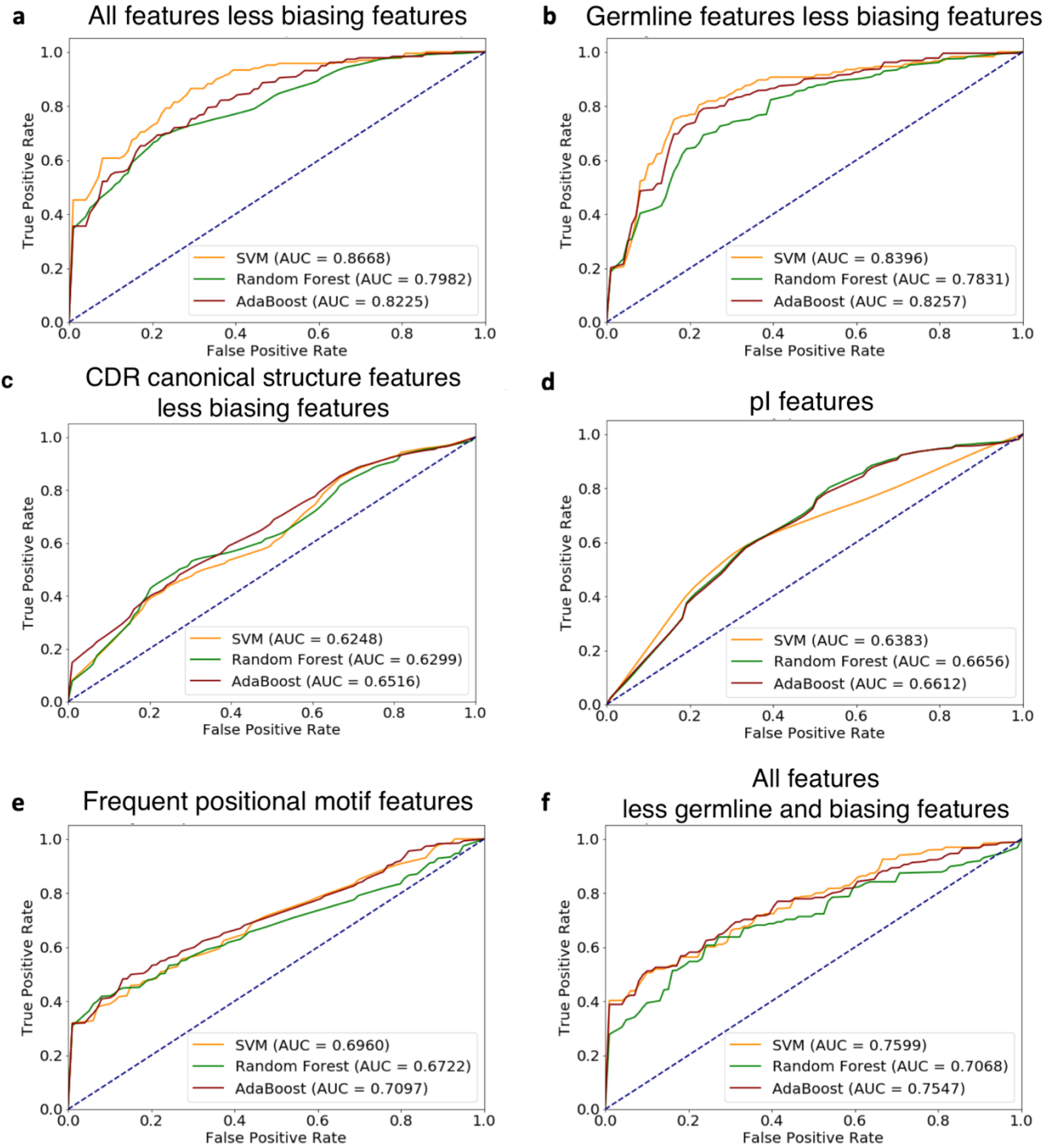

**Figure 3. Area Under ROC Curves (AUC) for classification using SVM, random forest AdaBoost algorithms, while excluding biasing features and their associated features.** (a) AUC based on all included features, (b) AUC based on germline features, (c) AUC based on CDR canonical structure features, (d) AUC based on pI features, (e) AUC based on frequent positional motifs features, (f) AUC based on all features excluding all germline features and associated CDR canonical structure features.

Some identified important features, such as Germline IGHJ4, had small percentages of differences of frequency between the MMP-IGHV-targeting and the IGHV-reference set when performing feature selection. It was possible to identify that such features as important because feature selection identifies combinations of important features and not independent important features.

**Contribution of features to classification for MMP-IGHV-targeting vs IGHV-reference sets**

Since no biasing features are identified in the MMP-IGHV-targeting and IGHV-reference sets, the full set of extracted features are used in classification (Figure S2). SVM yielded an AUC of 0.6941 (Figure S2a). Classification while retaining one feature-at-a-time (Figure S2b-e) yielded a range of AUC values: germline, AUC=0.5658; CDR-canonical structure, AUC=0.6023; pI, AUC=0.5648, and frequent positional motifs, AUC=0.6434. The AUCs when retaining one feature-at-a-time were lower than the AUCs when using all features. The classification yielded AUCs for CDR-canonical structure and frequent positional motifs that are higher than AUCs for germline and pI. These findings suggest that combinations of extracted features can better distinguish MMP-IGHV-targeting vs IGHV-reference set antibody sequences than a single type of feature.

**Comparing analyses on MMP-targeting vs PDB-reference sets and MMP-IGHV-targeting vs IGHV-reference sets**

The first case study, MMP-targeting vs PDB-reference sets, aimed to compare the MMP-targeting set against a diverse set of non-targeting sequences culled from the PDB. Frequency analysis identified biasing features and their associated features, which were removed for subsequent ASAP-SML analysis. To provide a more impartial analysis, we sought a second case study, where the reference set consisted of sequences with the HV germline that was predominant in the MMP-target set. Salient features identified for the two data sets were compared (Table S4a). When using FET, there were 7 common features out of 28 significant heavy-chain features identified in MMP-targeting vs PDB-reference analysis, and out of 17 significant heavy-chain features identified in MMP-IGHV-targeting vs IGHV-reference analysis. When using importance analysis, there were 12 common features out of the 36 important heavy chain features identified in MMP-targeting vs PDB-reference analysis and out of 19 important heavy chain features identified in MMP-IGHV-targeting vs IGHV-reference analysis. Four features were identified in common across FET and importance analysis (CDR-H3 canonical structure type 2, pI range 0.0-3.5, and motifs 2_YG and 5_YY), indicating that these features are consistently distinct when comparing the MMP sets against either of the two reference sets. However, this analysis alone cannot directly determine whether these features possess biological significance.

AUC plots (Figure 3 vs Figure S2) for the two case studies are compared. Overall, AUCs in the MMP-IGHV-targeting vs

IGHV-reference sets were lower than those in the MMP-targeting vs PDB-reference sets. In the germline case, mutually exclusive features contributed heavily to the classification of the MMP-targeting vs PDB-reference sets, while there were no such mutually exclusive features in the MMP-IGHV-targeting vs IGHV-reference sets. When classifying using only CDR canonical structures, classification accuracy was lower for the MMP-IGHV-targeting vs IGHV-reference sets than for the MMP-targeting vs PDB-reference sets. The latter analysis used features associated with both the heavy and light chains, while the former analysis used only the heavy-chain features. Classification using pI features yielded similar AUC results. The low frequency of positional motifs within MMP-IGHV-targeting and IGHV-reference sets contributed to low AUC result. Importantly, for both analyses, using combinations of features was more effective in classification than using any one single feature. Determining how these combinations affect biological function requires a more diverse and detailed experimental dataset than the one utilized in this study.

**Design recommendations using decision trees for the MMP-targeting vs PDB-reference sets**

To explore how the output of the ASAP-SML analysis could guide antibody design, a design recommendation tree (Figure 4) was constructed for the comparative data based on the corresponding decision tree (Figure S3). Each path from root to a blue node highlights combinations of feature values, which are either present or absent, that are more likely to be associated with the targeting antibody sequences based on the statistical analyses described above. Each combination provides a recommendation for feature values to include (those along true branches) and to exclude (those along false branches). Since some combinations of these features appear in MMP-targeting antibodies, decision trees may have value in designing collections of additional sequences in search of further function-disrupting sequences. Split efficiency, error rate, and the root-to-node path length are considered when identifying the best design recommendation or when analyzing design tradeoffs. As expected from identifying the IGHV3-23 germline as a biasing feature, utilizing the presence of feature-value SC1 (IGHV3-23 germline for HV) results in identifying 92.50% of MMP-targeting sequences, with a 5.67% error rate. Utilizing the presence of SC1 and absence of SC2 (IGLJ3 germline for LJ) results in identifying 78.75% of sequences with 4.04% error rate. The lowest error rate for the targeting set is the result of a combination consisting of the presence of SC1 (IGHV3-23 germline for HV), absence of SC2 (IGLJ3 germline for LJ), absence of SC5 (IGKJ1 germline for LJ), and absence of SC8 (IGLJ1 germline for LI) and absence of SC10 (PI 3.9375-4.375) resulting in a node with a 0% error rate for 49.38% of the MMP-targeting set. Design variations based on decision recommendation (including SC1 while avoiding other conditions) may lead to additional testing candidates. We note however that these recommendations are specifically based on our datasets and are only provided to depict the decision tree functionality of ASAP-SML.

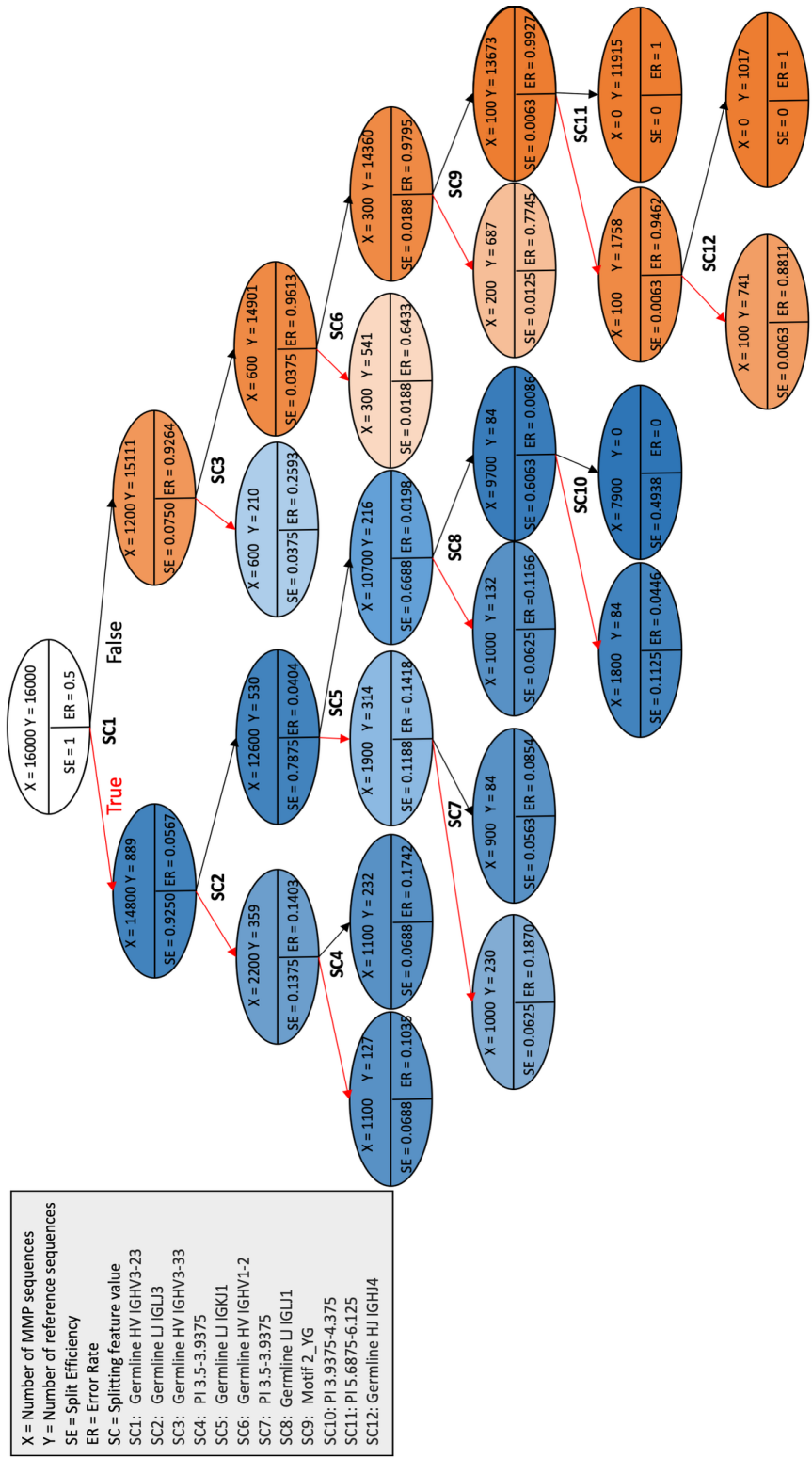

**Figure 4. Design recommendation tree for the MMP-targeting antibody test case.** Each node lists the number of MMP sequences (X), and the number of reference sequences (Y), along with the splitting efficiency and error rate. The label under each node, when present, reflects the splitting feature value and is expanded in the legend.

# Discussion

This paper describes the implementation of ASAP-SML, a pipeline for identifying features common in one set of antibodies in reference to another set. This pipeline extracts residue-based and CDR-H3 region features from primary amino acid sequences and supports several analyses to identify features and feature values that are significantly overrepresented in a targeting antibody set when compared to a reference set. Further, ASAP-SML builds a design recommendation tree to aid in identifying and evaluating combinations of feature values for inclusion or exclusion when designing further candidate targeting antibody sequences. We are not aware of any other analysis pipelines that analyze antibody sequences in the ways described within. As with all data-driven approaches, however, the value of the analyses and generalizability of the findings depend heavily on the availability of sufficient quantities of high-quality data.

In analyzing sets of MMP-targeting antibody sequences against PDB-reference sequences, we found that features associated with the antibody heavy chain are much more likely to differentiate our MMP-targeting sequences from the selected PDB-reference antibody sequences. This result is consistent with experimental findings that show that antibody heavy chains play dominant roles in antigen recognition (19, 42). Comparing the MMP-IGHV-targeting set against the IGHV-reference antibody set, ASAP-SML identified several salient features that were in agreement with those identified when analyzing the MMP-targeting vs PDB-reference sets. While we utilized sequence clustering to minimize redundancy and identified correlation between variables, we note that biases in the available datasets may explain these specific observations. The use of the pipeline enabled identification of some frequently occurring dipeptide motifs; however, the presence of such motifs does not necessarily imply any functional consequence. Importantly, the analysis shows that the ASAP-SML pipeline is capable of identifying salient features between targeting and reference sequence sets.

We developed design recommendation trees to identify combinations of feature values that can be used to generate additional sequences with features that distinguish between targeting and reference datasets. We expect that identified features and combinations thereof will be useful for the purpose of augmenting existing antibody libraries by identifying related sequences that have a higher probability of yielding antibodies that inhibit their targets, or for enhancing the affinities of existing targeting antibodies through affinity maturation. Further, for a sufficiently diverse targeting data set, identified features can be incorporated into various antibody computational synthesis approaches including de novo design or the redesign of existing antibodies (52). Expanding the ASAP-SML pipeline to include properties of targeted epitopes (if known), antibody subtypes, and/or more properties that can be determined via computational antibody prediction (6, 53) and utilizing this approach in combination with experimental data will enable further refinements to the pipeline.

While we evaluated the utility of ASAP-SML for an MMP-targeting set, we expect that ASAP-SML will be utilized as a general analysis pipeline for the identification of antibody features that alter the biological functions of their targets, conditional on the availability of datasets that support such analyses. This includes both antibodies targeting enzymes from other protein families and antibodies that disrupt additional biological processes such as viral entry. The ASAP-SML approach should be compatible with any antibody discovery effort as long as a diverse and representative set of sequencing data is available for both targeting and reference sets.

## Supporting information

**Figure S1. Contribution of features to classification for MMP-targeting vs PDB-reference set.** AUC for classification using SVM, random forest AdaBoost algorithms when (a) using all features, (b) retaining germline features, (c) retaining CDR canonical structure features, (d) retaining pI features, and (e) retaining frequent positional motif features, (f) excluding germline features.

**Figure S2. Contribution of features to classification for MMP-IGHV-targeting vs IGHV-reference set.** AUC data is reported for classification when (a) using all features, (b) retaining germline features, (c) retaining CDR canonical structure features, (d) retaining pI features, and (e) retaining frequent positional motif features.

**Figure S3. Decision tree output during design recommendation step when analyzing the MMP-targeting set vs PDB-reference set**. With a desired dataset size of 160, which is the size of representative MMP-targeting sequences, and $k=100$ sampling iterations, each set had 160*100 sequences. The label within each node reflects the following: the feature value, the Gini impurity score, the number of samples within the tree rooted at that node, a value providing a listing of the number of samples that are from the reference set followed by the number of samples that are from the MMP-targeting set, and a node classification label indicating if the node is dominated by reference or MMP-targeting sequences.

**Table S1. Detailed data for the collected MMP-targeting antibody sequences.** (a) original sequences, (b) extracted features, (c) representative MMP-targeting set sequence IDs after BLASTCLUST and corresponding sequences in the original set, (d) representative MMP-IGHV-targeting set heavy chain sequence IDs after BLASTCLUST and corresponding sequences in the original set.

**Table S2. Detailed data for representative sequences in MMP-targeting vs PDB-reference sets.** (a) sequences for MMP-targeting set, (b) extracted features for MMP-targeting set, (c) sequences for PDB-reference set, (d)

extracted features for PDB-reference set, (e) distribution of features, (f) statistical testing and feature selection scores for features in MMP-targeting and PDB-reference sets, (g) Jaccard coefficient association scores for features within the MMP-targeting set and within the PDB-reference set.

**Table S3. Detailed data for representative sequences in the MMP-IGHV-targeting and IGHV-reference sets.** (a) sequences for MMP-IGHV-targeting set, (b) extracted features for MMP-IGHV-targeting set, (c) sequences for IGHV-reference set, (d) extracted features for IGHV-reference set, (e) distribution of features, (f) statistical testing and feature selection scores in the MMP-IGHV-targeting and IGHV-reference sets, (g) Jaccard coefficient association scores for features within the MMP-IGHV-targeting set and within IGHV-reference set.

**Table S4. Comparison of salient features for the two comparative sets: the MMP-targeting vs PDB-reference set**s **and the MMP-IGHV-targeting vs IGHV-reference sets.**

# Acknowledgements

We appreciate the contribution of Niroshan Anadasivam, funded under NSF REU award 1560388, in running the isoelectric point calculations for the CDR-H3 region.